# Cross-point architecture for spin transfer torque magnetic random access memory

W.S. Zhao *Member, IEEE*, S. Chaudhuri, C. Accoto, J-O. Klein *Member, IEEE*, C. Chappert *Member, IEEE,* and P. Mazoyer

*Abstract*—Spin transfer torque magnetic random access memory (STT-MRAM) is considered as one of the most promising candidates to build up a true universal memory thanks to its fast write/read speed, infinite endurance and non-volatility. However the conventional access architecture based on 1 transistor + 1 memory cell limits its storage density as the selection transistor should be large enough to ensure the write current higher than the critical current for the STT operation. This paper describes a design of cross-point architecture for STT-MRAM. The mean area per word corresponds to only two transistors, which are shared by a number of bits (e.g. 64). This leads to significant improvement of data density (e.g. 1.75 $F^2$/bit). Special techniques are also presented to address the sneak currents and low speed issues of conventional cross-point architecture, which are difficult to surmount and few efficient design solutions have been reported in the literature. By using a STT-MRAM SPICE model including precise experimental parameters and STMicroelectronics 65 nm technology, some chip characteristic results such as cell area, data access speed and power have been calculated or simulated to demonstrate the expected performances of this new memory architecture.

## I. Introduction

Magnetic random access memories (MRAM) promises stable non-volatility, fast write/read access speed and infinite endurance etc. It attracts considerable research effort from both academics and industries since 2000 [1-4]. The first MRAM chip based on the field induced magnetic switching (FIMS) approach was commercialized in 2006 [5] and today it is primarily used in the fields of aerospace and aeronautics thanks to its radiation hardness. The wide application of MRAM is mainly limited by FIMS, which requires too high currents (e.g. ~10mA) for storage cell state changing [6]. New switching approaches like thermally assisted switching and domain wall motion are currently under intense R&D [2, 7-8].

Spin transfer torque (STT) is one of the most promising switching approaches thanks to its high power efficiency and fast writing speed [3-4, 9]. As the magnetic tunnel junction (MTJ) nanopillar or MRAM storage element is smaller than 100nm, a low spin-polarized current (<200 µA@65 nm node) can switch its state (see Fig.1). An MTJ nanopillar is mainly composed of three thin films: a thin oxide barrier and two ferromagnetic (FM) layers. In standard applications, the magnetization of one FM layer is pinned, while the other is free to take the two orientations, Parallel (P) or Anti-Parallel (AP) corresponding to resistances $R_P$ and $R_{AP}$. The Magnetoresistance TMR= $(R_{AP}-R_P)/R_P$ characterizes the amplitude of this resistance change and it rises up to 200% with MgO barrier [10]. STT approach opens the door to build up the first true universal memory with MRAM, which should provide both large capacity (> Gigabit) and high speed (<ns). Beyond the memory applications, new non-volatile logic circuits can also be expected [11-12], which would have more important impact on the current computing systems suffering hardly from high power issues.

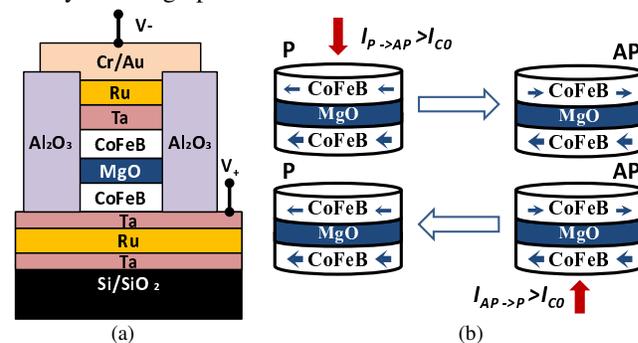

Fig. 1. (a) Vertical structure of an MTJ nanopillar composed of CoFeB/MgO/CoFeB thin films. (b) Spin transfer torque switching mechanism: the MTJ state changes from parallel (P) to anti-parallel (AP) as the positive direction current $I_{P->AP}$>$I_{C0}$, on the contrary, its state will return to P state with the negative direction current $I_{AP->P}$>$I_{C0.}$

Since 2007, a number of STT-MRAM prototypes have been demonstrated based on advanced technology nodes [3-4, 13-14]. However the cell size is always very large, far from that of flash memory and Dynamic RAM (DRAM) [15]. This limits its potential capacity. For instance, the cell size is 0.3584 um$^2$, more than 80 $F^2$ (F: Feature) in the last 64Mb prototype [13]. This is firstly caused by the high threshold current $I_{C0}$ (see Fig.1b), depending mainly on the thickness and material of free layer (e.g. $Co_{60}Fe_{20}B_{20}$) [16]. The second reason is the MTJ access architecture, which is based on 1 or 2 selection transistors associated with one MTJ [2-4, 13-14] (see Fig.2a). This design allows the STT-MRAM to benefit directly from all the peripheral address and control circuits of DRAM, but it leads to large cell size due to the selection transistor, and the density of STT-MRAM depends only on the CMOS part. Fig.2b shows the 65 nm layout implementation of the conventional STT-MRAM access design. To ensure the high switching speed (e.g. 10 ns), 100 µA is needed to pass through the selection transistor [17], which should be thus large enough, e.g. W=0.4 µm instead of the minimum width. In this case, the cell size becomes at least 56 $F^2$. Moreover the high number of contacts between MTJ and CMOS raises the fabrication difficulty and cost.

*Cross-point* or *crossbar* architectures were proposed to relax the density limitation of two terminal memristive devices imposed by the CMOS circuits [18-21] (see Fig.3). However, they suffer from either the *sneak currents* or low data access speed, which are difficult to surmount and few efficient design solutions addressing this issue have been reported previously in the literature. In this paper, we present a new STT-MRAM access design based on *cross point* architecture to overcome these drawbacks while keeping high storage density. By using an accurate STT-MTJ compact model [22] and CMOS 65 nm design kit [23], hybrid simulations and theoretical calculations have been performed to analyze the storage density and data access speed of this cross point STT-MRAM architecture.

The rest of the paper is organized as follows: section 2 describes the scheme of cross point architecture and peripheral write/read circuits. Some special design techniques to avoid the sneak currents and accelerate the data access speed are also introduced; in sections 3 and 4, we present the transient simulation and performance analysis of the new architecture; some conclusions are addressed in the last section.

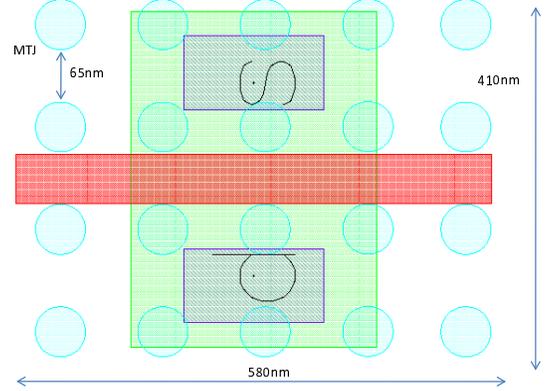

Fig. 3. The layout implementation promises the best area efficiency, where the die area per storage bit is $F^2$ and the selection transistor is shared by a number of MTJs associated in the same word (e.g. 8).

## II. CROSS POINT ARCHITECTURE SCHEME AND ITS OPERATIONS

### A. Global structure of new MTJ access design

The cross point architecture of STT-MRAM is composed of four parts: MTJ array for data storage, MTJ array for reference, selection transistors, write and read circuits (see Fig.4). The word lines and bit lines address respectively the two terminals of MTJ in their cross-point as shown in the inset. In this design, the storage density is determined by the minimum distance between two MTJs instead of the CMOS part for the conventional one. There are two selection transistors (one NMOS and one PMOS) per word composed of *N* bits (e.g. 4). Two lines of MTJ array (brown) are dedicated for data reference, which reads respectively the odd and even words (green) by the sense amplifier (S.A) [24]. There is one write/read circuit per bit. Besides the extremely small cell area potential of this cross point architecture, the implementation of MTJs could become much easier as there are much fewer contacts between MTJs and CMOS circuits and they are used only in the edge of the MTJ array.

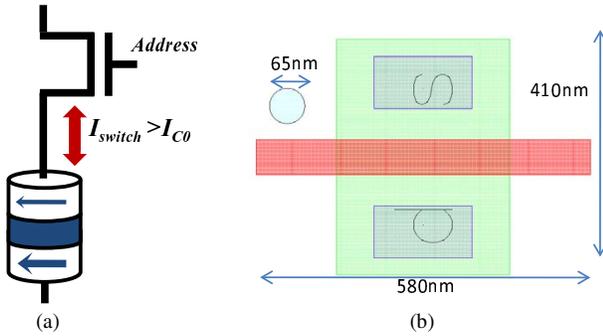

Fig. 2. Conventional STT-MRAM selection approach based on 1 Transistor and 1 MTJ. As $I_{write}$ passing through the MTJ and transistor should be higher than $I_{C0}$, a large selection transistor is required (a) Circuit diagram (b) Layout implementation; the size of selection transistor is about 56 $F^2$.

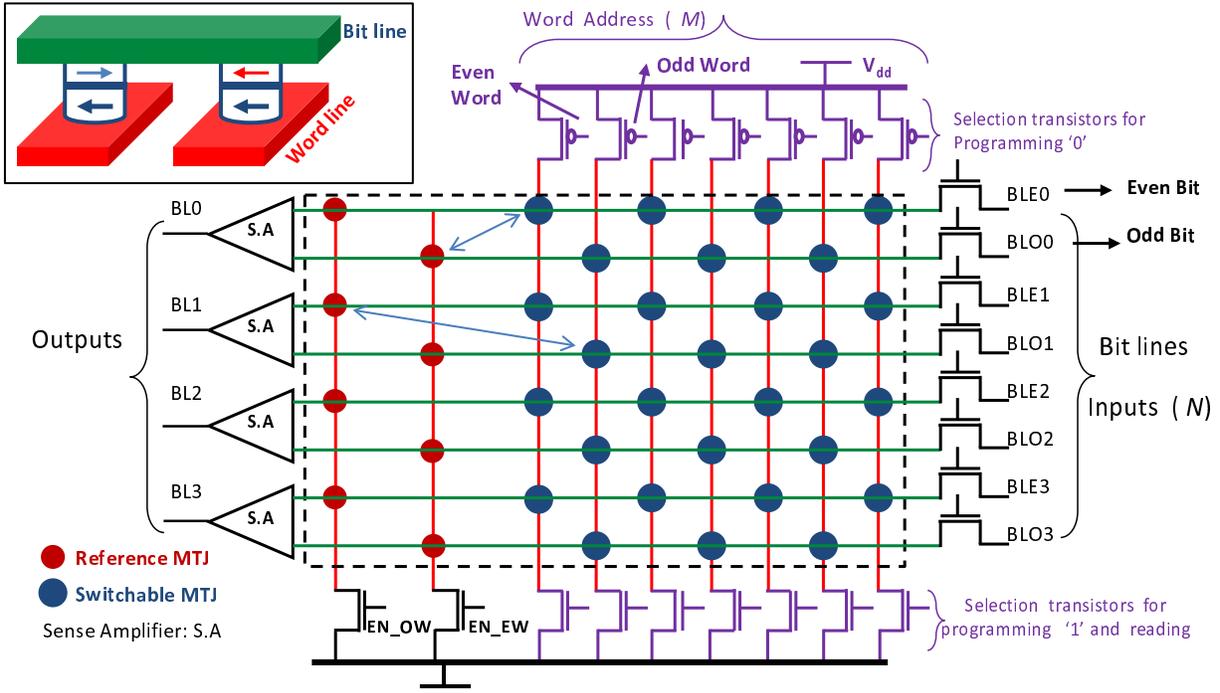

Figure. 4: Proposed Cross Point architecture for STT-MRAM (4x8 array). It includes four parts, a cross-point array of MTJs for data storage, a cross-point array of reference MTJ, write (right side) and read circuits (left side), the word selection circuits.

In the following subsections, the operating mechanisms of this cross point STT-MRAM will be detailed.

### B. Switching mechanism

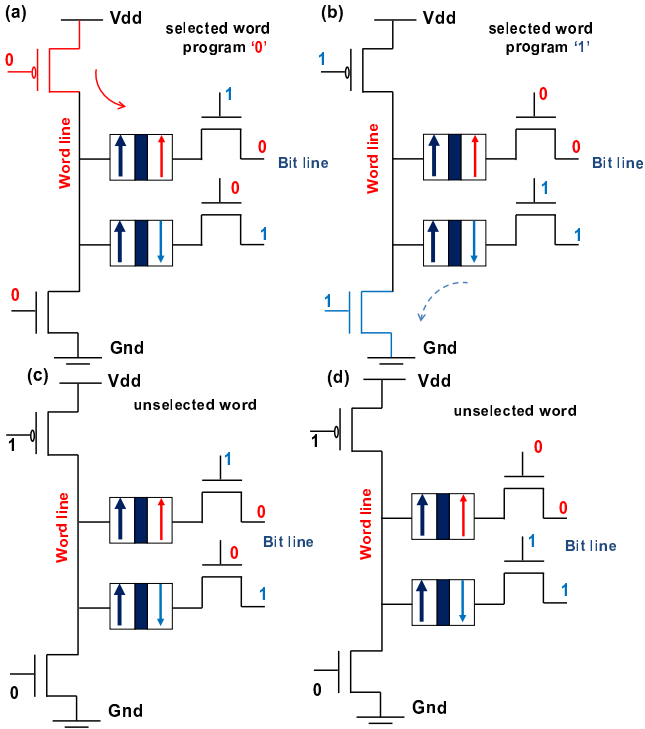

Fig. 5. Switching mechanism of MTJ in the cross point array: (a) For the selected word, the PMOS is active to program the selected bit to '0'. (b) For the selected word, the NMOS is active to program the selected bit to '1'. (c) and (d) For the unselected words, both the NMOS and PMOS are always inactive whatever the address and the value of bit lines.

As shown in Fig.1b and Fig.2a, the switching current of MTJ is bidirectional. Thus the word selection transistors should be able to generate the currents in two directions according to different values sent by the bit lines. Fig.5 shows an example of word selection with two transistors (one NMOS and one PMOS) per word. The word lines and bit lines connect respectively the reference layer and free layer terminals of MTJ as shown in the inset of Fig.4. For the selected words, each time there is only one active transistor to program '0' and '1'. For the unselected word, both transistors are always inactive whatever be the value of bit lines. This allows the selected word to be addressed with minimum noise (i.e. $I_{off}$ of selection transistors). In the selection word, the selected bit should also be well addressed as the current passing through the selection transistors is limited by their width. Series MTJ programming in the same word allows the best area efficiency as the selection transistors can be achieved with small width. We assume that the minimum area of selection transistors able to pass through a write current $I_{wite} > I_{C0}$ and ensure a switching duration <=10 ns is 56 $F^2$ (see also Fig.2). The minimum cell area per bit is then 112/N $F^2$, in the cross-point array, where *N* is the number of bits or MTJs in the same word. As the bit lines including both write /read circuits can be shared by the words, they can be negligible for the cross-point array with high number of words (e.g. 1024).

Parallel MTJ programming in the same word allows the best access speed, as shown in Fig.5a and Fig.5b, the programming of one word can be operated in the phases '0' and '1'. The program duration of one word is 2× *τ*, where *τ* is the switching delay for single MTJ cell [17]. This especially high access speed makes this architecture suitable for embedded solutions. However the parallel MTJ switching needs much larger die area than that of sequential switching as the selection transistor should be large enough to ensure the necessary current higher than $N \times I_{C0}$ and the area of one storage bit is ~56 $F^2$. This large area limits its interest to build up standalone memories like the 1 T+1 MTJ structure.

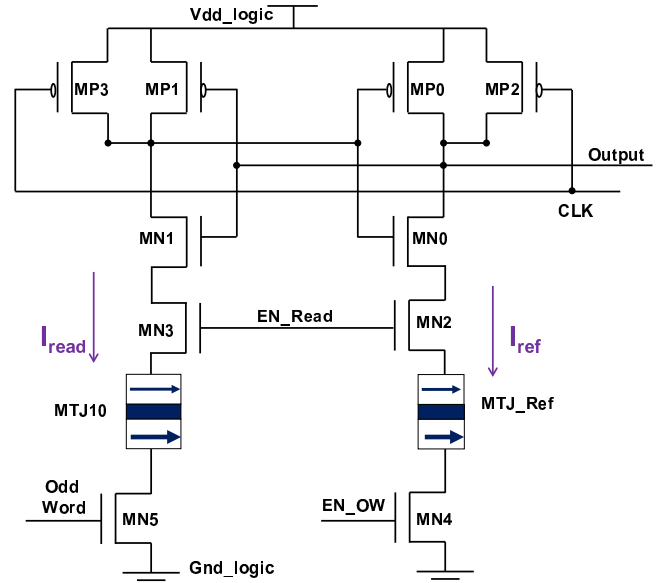

Fig. 6. Pre-Charged Sense Amplifier (PCSA) for data sensing: MP0-3 and MN0-2 constitute the amplifier; MN4 and MN5 serves as respectively the reference and word selection; MN2 and MN3 plays the role of "Enable".

### C. Reading mechanism

The read operation of data stored in MTJ is currently one of the major challenges for MRAM R&D as the TMR ratio is limited to 200% or 300% [10, 15]. This limits the sensing margin between logical state '0' and '1'. Moreover the resistance property of MTJ is very sensitive to deposition process variation. A sense amplifier performing with high reliability is then required. Fig.6 shows a pre-charge based sense amplifier (PCSA), which has demonstrated the best tolerance to different sources of variation [24], while keeping high speed and low power. In this SA, the circuit is first pre-charged with "CLK"= '0'. The data stored in MTJ state (AP or P) can be evaluated to logic level at the "Output" as "CLK" is changed to '1'. In order to obtain the best sensing margin between $I_{ref}$ and $I_{read}$, the resistance value of $R_{ref}$ should be equal to $(R_{AP}+R_P)/2$. This can be achieved by changing the surface of MTJ according to Eq.1 [25].

$$R_P = \frac{tox}{k \times \overline{\varphi}^{-1/2} \times Surface} \times \exp(1.025 \times tox \times \overline{\varphi}^{-1/2}) \quad (1)$$

where $\overline{\varphi} = 0.4$ is the potential barrier height of crystalline MgO [10], *tox* is the thickness of oxide barrier and *surface* is the MTJ area. *k* is a factor calculated from the resistance-area product (R.A) value of MTJ, which depends on the material composition of the three thin layers. For the MTJ nanopillar, we used the values with R.A=10 $\Omega\mu m^2$, $k = 332.2$.

One of the most important drawbacks for cross-point architecture is the *sneak paths or sneak currents* (see Fig.7) [18], which introduces significant perturbation for data sensing due to the small margin between $I_{ref}$ and $I_{read}$. To avoid the influence of sneak currents, two special design considerations have been developed in this cross-point architecture. The first one is the balanced sensing structure where there is the same number of MTJ in both branches of sense amplifier (see Fig.4). There are one reference MTJ and *M* MTJs for storage in each side of sense amplifier, where *M* is the number of words in the cross-point STT-MRAM array. The odd and even words are read respectively with the different reference MTJ, which are associated to two branches of the sense amplifier. This design allows the disturbance of the sneak currents from the same bit address to be mitigated during data sensing (see Eq.2-3).

$$I_{ref\_final} = I_{ref} + \sum_{i=0}^{M/2} I_{sneak\_i} \quad (2)$$

$$I_{read\_final} = I_{read} + \sum_{i=0}^{M/2} I_{sneak\_i} \quad (3)$$

where $I_{ref\_final}$ is the current passing through *the selection transistor* of reference MTJ and $I_{read\_final}$ is the current passing through *the selection transistor* of MTJ for storage. $I_{ref}$ is the current passing through the reference MTJ and $I_{read}$ is the current passing through the MTJ for storage (e.g. $I_{01}$ in Fig.6). $I_{sneak\_i}$ are the parasitic currents in the sneak paths at other words with the same parity. The MTJs connecting to the two branches of sense amplifier are distributed into different parity words (see Fig.4 & 7). This allows a single read operation to access each word with the sacrifice of capacity as only a half of the cross-points are occupied. We can use all the cross-points to obtain the maximum density, however one word should be accessed with two cycles of data reading.

The second design consideration to overcome the sneak currents is to implement parallel data reading, as shown in Fig.4 and Fig.7-8. *N* sense amplifiers are used to detect the data in parallel. For each MTJ associated in the same word, parallel reading allows them to avoid the sneak currents from the other word as there aren't any floating nodes in the bit lines (see Fig.8b). This allows important power saving compared with series sensing bit by bit (see Fig.8a) and the detailed power calculation will be presented in the subsection IV.B. The bit lines are all pre-charged to $V_{read}$ before data evaluation (see also Fig.6) and the selected word line is always grounded during the sensing.

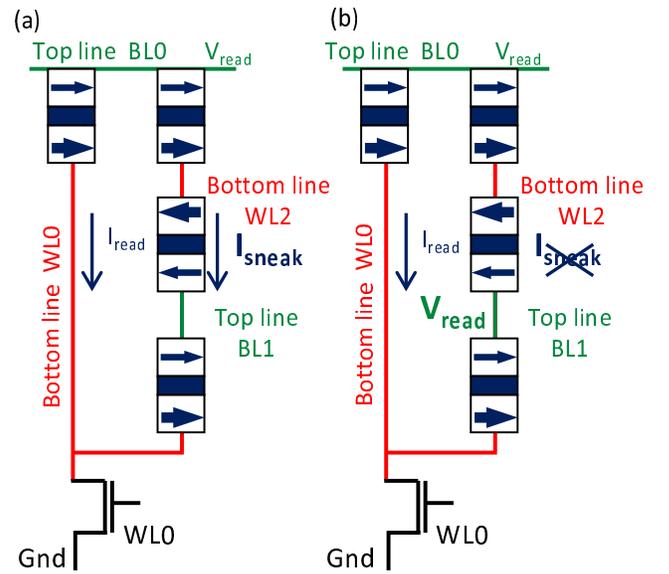

Fig. 8. (a) Series bit sensing suffers from the sneak currents as the nearby bit lines (e.g. BL0 and BL1) are "floating". (b) Parallel reading avoids floating nodes in the bit lines and then allows the sneak currents from the other word address to be eliminated.

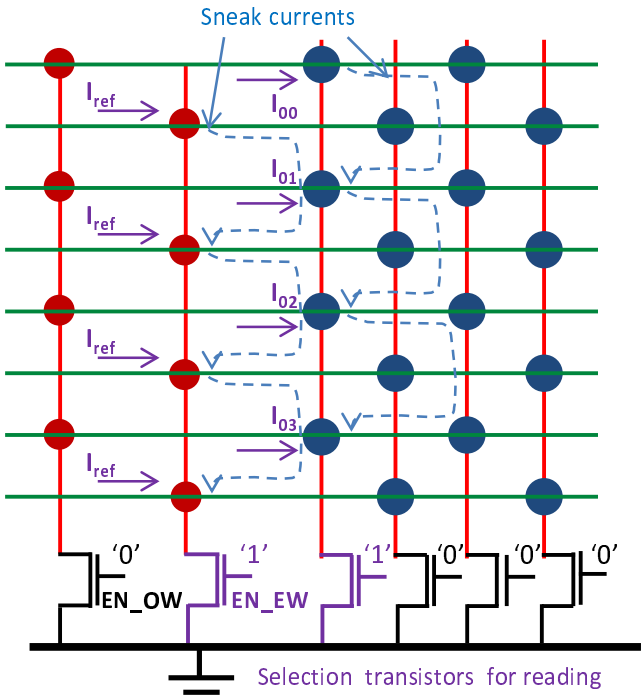

Fig. 7. Cross-Point architecture suffers from sneak currents, which disturb data sensing. Balanced architecture of sense amplifier and parallel reading approach allows the sneak currents to be mitigated.

With combined use of these two design techniques, noise influence of sneak currents can be neglected for data sensing. It is noteworthy that achieving correct write operations is a big challenge for cross-point RRAM and PCRAM [15], as the sneak paths could be much lower resistive than the addressed cell due to the high $R_{ON}/R_{OFF}$ ratio (i.e. 100). The main part of writing current generated by the voltage difference between word line and bit line pass through the sneak paths. This leads unwanted write operation, particularly the cells in $R_{OFF}$ state. Nevertheless, the sneak currents do not affect the write operation of cross-point architecture for STT-MRAM, as the $R_{AP}/R_P$ ratio is relatively low (i.e. 1.5) and the addressed cell presents always the lowest resistance compared to the sneak paths. For instance, if the addressed cell is in $R_{AP}$ state, the minimum sneak path resistance is $3 \times R_P$ (see Fig.8b). $R_{AP} < 3 \times R_P$, if TMR ratio is lower than 300%. The same as cross-point R-RAM and PCTAM, the sneak currents will drive additional power for the write operations and this point will be detailed in sub-section IV.B.

Another advantage of parallel sensing is to improve greatly the access speed, which is another limitation of conventional cross-point architecture. In general, non-volatile memory is accessed more frequently for reading than programming. For instance, in the normally off system [26]. Series programming and parallel sensing could be thus the best tradeoff between area and speed performance for cross-point STT-MRAM.

### III. SIMULATION OF 4×4 CROSS-POINT ARRAY

By using a CoFeB/MgO/CoFeB STT-MTJ compact model [18] and STMicroelectronics CMOS 65 nm design-kit [19]; mixed MTJ/CMOS simulations have been performed to demonstrate the write and read operations of this cross-point architecture. This model has been developed based on the physical theories and experimental measurements of perpendicular magnetic anisotropy (PMA) MTJ [17, 27]. The shape of MTJ is circular with diameter=65 nm. The other main device parameters are shown in Table.I.

TABLE I
PARAMETERS USED IN STT-MTJ COMPACT MODEL

| Parameter | Description | Default Value |
|---|---|---|
| $H_K$ | Anisotropy field | $113.0 \times 10^3$ A/m |
| $M_S$ | Saturation magnetization | $456.0 \times 10^3$ A/m |
| $a$ | Magnetic damping constant | 0.027 |
| RA | Resistance.Area product | 10 ohm/um$^2$ |
| $J_C$ | Critical current density | $5.7 \times 10^6$ A/cm$^2$ |
| $t_{CoFeB}$ | Free layer height | 1.3 nm |
| $t_{MgO}$ | Oxide layer height | 0.95 nm |
| T | Temperature | 300 K |
| surface | MTJ surface | 65 nm x 65 nm x $\pi/4$ |
| TMR | TMR ratio | 150% |
| V | Volume of free layer | surface x1.3 nm |

Fig.9 presents the target configuration of 4×4 cross-point STT-MRAM; BL0-3 and WL0-3 represent respectively the bit line and word line address. Fig. 10 (a) and (b) demonstrate the mixed simulation of parallel writing /reading for this 4×4 cross-point STT-MRAM and confirm the expected operations shown in the section II. For instance, it takes only one cycle of switching duration, ~1.1 ns driven by the signal "EN_Write" to program a word to "0000" or "1111". Thanks to the fast computing speed of the PMA MTJ compact model, the simulation of this 4×4 cross-point memory can be performed in ~30 minutes in a medium performance CAD server (two Xeon: 4-Core, 12MB cache, 2.4GHz and 8GB 1.3GHz RAM).

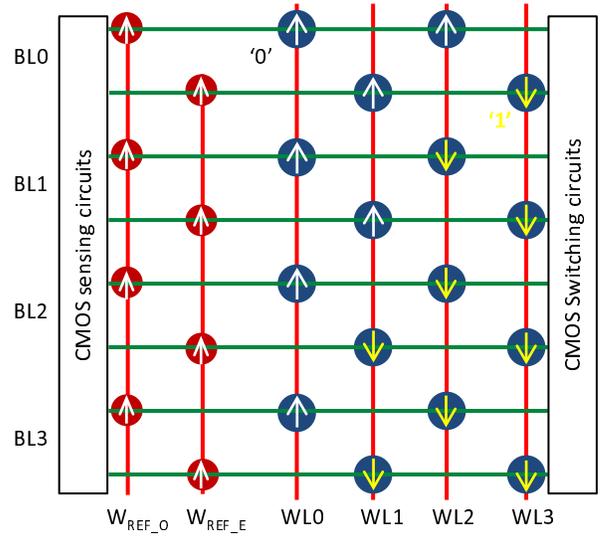

Fig. 9. Target 4×4 cross-point STT-MRAM configuration for the simulation.

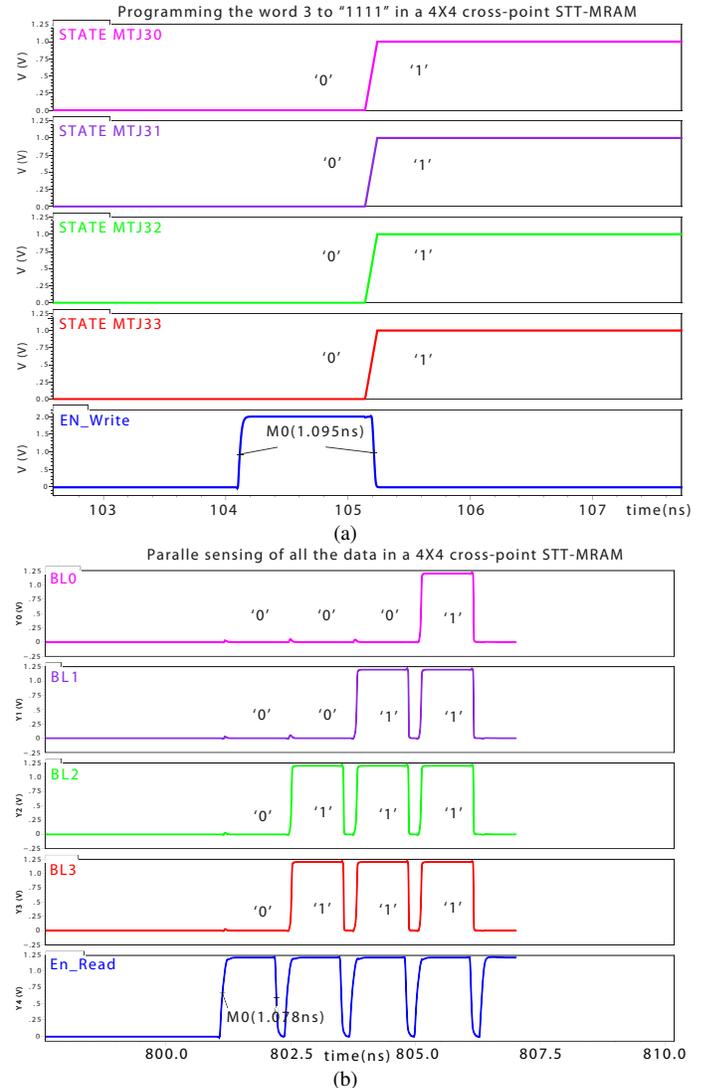

Fig. 10. (a) Parallel programming of the word 'WL3' to "1111" within ~1.1ns. (b) Parallel sensing of a 4×4 cross-point array within ~5ns (~1.2ns/word).

In order to keep the same data read access speed with that of data programming, the pulse duration of "En_Read" (see also Fig.6) is set to ~1.1ns. The word address changes between two "En_Read" pulses during ~100 ps and the data stored in this 4×4 cross-point STT-MRAM can be detected word by word in ~5ns. It is noteworthy that the sensing speed can be accelerated up to ~200 ps/word [20], which would lead to an asymmetric delay between the programming and reading operations. Nevertheless, this asymmetric delay is nearly ubiquitous in non-volatile memories and it may present some advantages in terms of power and access speed as the non-volatile memories are read more frequently than programmed.

## IV. PERFORMANCE ANALYSIS

Based on the mixed MTJ/CMOS simulation and theoretical calculation, we analyze the cell area density, access speed and power consumption of cross-point STT-MRAM in this section. This would help designers to obtain the best performance tradeoff towards different applications.

### A. Cell area and data access speed

As shown in Fig.4, a cross-point STT-MRAM array is composed of three CMOS parts: bit write circuits, bit read circuits and word selection circuits. In order to compare the area efficiency with the conventional 1 T+ 1 MTJ structure, only the CMOS footprint is considered for the mean cell area $A_{CP}$ calculation, which is described by the following equation:

$$A_{CP} = \frac{N \times A_{SA} + N \times A_{Write} + (M+2) \times A_{Se}}{N \times M} \quad (4)$$

where $A_{SA}$ is the area of a sense amplifier = ~40 $F^2$ (see Fig.6), $A_{write}$ is the area of write circuit per bit =~112 $F^2$, $A_{Se}$ is the area of selection circuit per word =~112 $F^2$, $N$ and $M$ are respectively the number of bits per word and number of words per array. There are 2 words per array for reference.

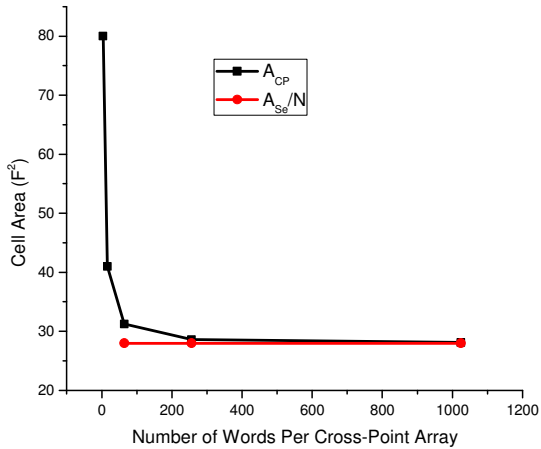

Fig. 11. As N=4, cell area analysis of cross-point STT-MRAM with different number of words per array. The write and read circuits of cross-point area can be ignorable as M>>N.

If $M$ is much bigger than $N$, Eq. 4 can be simplified to Eq.5, which means that the die area of write and read circuits can be nearly neglected to calculate the cell area density. Fig.11 confirms clearly this conclusion. For instance, if $N$=4 and $M$=1024, $A_{CP}$ = ~28.14 $F^2$ by Eq.4 and 28 $F^2$ by Eq.5.

$$A_{CP}=A_{Se}/N \quad (5)$$

According to Eq.5, the most efficient method to reduce the cell area is to increase $N$. Fig. 12 shows that $A_{CP}$ can be theoretically reduced down to 3.5 $F^2$ and 1.75 $F^2$ with N= 32 and 64 respectively, where F is the feature size for CMOS circuits. They are inferior to the practical minimum cell area of cross-point memory, 4 $F_M^2$, where $F_M$ is the feature size of STT-MRAM. If $F_M$ =F, $A_{CP}$ depends on MTJ layout instead of CMOS circuits for the 1 T + 1 MTJ structure. If $F_M$ becomes very small and independent from that of CMOS, the theoretical minimum area < 1.75 $F^2$ can be achieved. For instance, if CMOS and STT-MRAM are based on respectively 65nm (F) and 40nm ($F_M$), the practical minimum cell area, 4 $F_M^2$, can be down to 1.51 $F^2$. This extremely small cell area allows >Gb non-volatile storage to be implemented in 100 $mm^2$ die size and promises great potential for standalone memory applications.

Bigger $N$ leads to an unwanted linear increase in word programming duration, as shown in Fig.12. New STT-MRAM switching circuits can potentially manage this performance degradation. For instance, the activation of switching enable signal "EN_Write" results from the comparison between the target configuration and previous storage in order to eliminate useless write operations [28]. It is important to underline that the word reading keeps the same speed whatever be the number of bits per word thanks to the parallel sensing approach (see Fig.10).

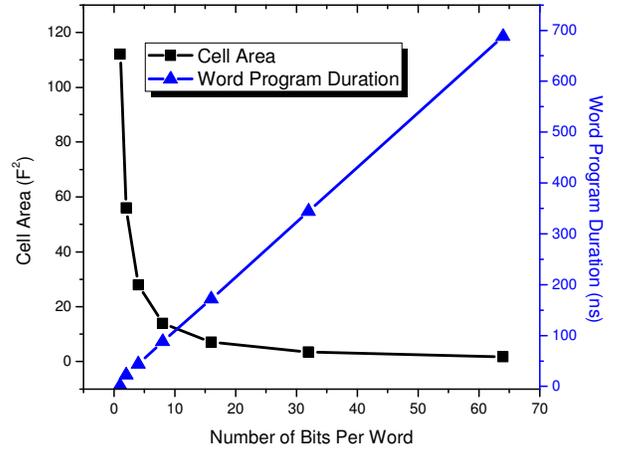

Fig. 12. For M=1024, area and speed analysis of cross-point STT-MRAM with different number of bits or MTJs per word. 1.75 $F^2$ cell area can be achieved as the number of bits per word=64.

As mentioned in subsection II.B, word-programming duration of this cross-point memory can be a constant (i.e. 2 cycles of MTJ switching duration $\tau$) with the cost of large area. This high speed allows STT-MRAM to be used in register, Flip-Flop, cache memory and logic circuits. This degree of freedom in the tradeoff between data access speed and storage density makes cross-point STT-MRAM to be used as unique technology or universal memory in the complex memory hierarchy. Note that the switching duration $\tau$ can be adjusted by changing $I_{write}$ according to the dynamic switching theory of spin transfer torque, described in Eq.6 [17]. For instance, $\tau$ equals to ~10 ns and ~1.1 ns if $I_{write}$ is designed to $1.3 \times I_{C0}$ and $3 \times I_{C0}$ (see also Fig.10). In these two cases, the area of selection transistors $A_S = $ ~112 $F^2$ and ~405 $F^2$ respectively to let the $I_{write}$ pass through MTJ, The latter presents a cell area $A_{CP} = $ ~12.6 $F^2$ if $N= 32$.

$$\frac{1}{<\tau>} = [\frac{2}{C + \ln(\frac{\pi^2 \xi}{4})}] \frac{\mu_B P}{em(1+P^2)} (I_{write} - I_{c0}) \quad (6)$$

where $C \approx 0.577$ is the Euler's constant, $\xi = E/k_B T$ the activation energy in units of $k_B T$, $k_B$ the Boltzmann constant, $T$ the temperature, $P$ the tunneling spin polarizations of the ferromagnetic layers, $\mu_B$ the Bohr magneton, $e$ the elementary charge and $m$ is the magnetic moment of free layer.

### B. Power Consumption

The total power of modern digital IC consists of dynamic and static dissipation. The latter presents currently ~40% of the total power and it will be increased up to ~60% with shrinking of the feature sizes [29]. The fast data access speed of this cross-point STT-MRAM allows it to be powered off completely in "idle" state and restarted on instantly. Thereby the first power gain of this new architecture design is the nearly zero static dissipation.

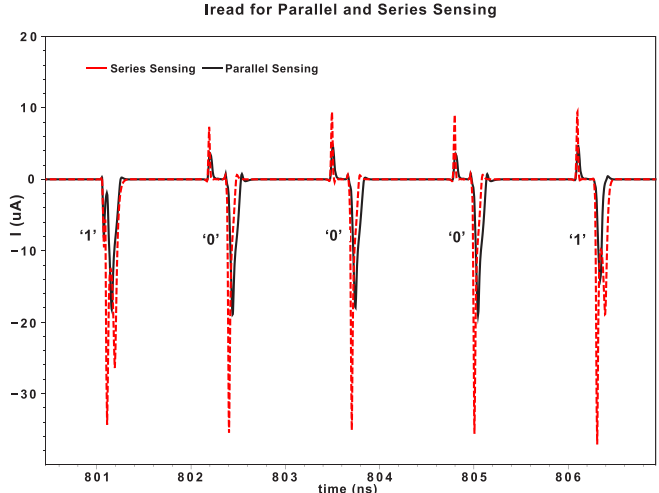

Fig. 13. $I_{read}$ of one sense amplifier (see Fig.6) or one bit line address for the parallel sensing and series sensing approaches.

$$P_{dynamic} = f_{data} \times \int_0^{Td} V_{dd} \times I_d(t) dt \quad (7)$$

where $f_{data}$ is the data throughput frequency, $Td$ the read or program duration, $I_d$ the current and $Vdd$ is the power supply.

The dynamic power for data reading can also be reduced compared with conventional cross-point architectures. As mentioned in subsection II.C, the parallel sensing approach for each word avoids the sneak currents, and then lowers the MTJ read current $I_{read}$. This is confirmed by the transient simulation shown in Fig.13. The MTJs associated with the same bit line address like BL0-3 in Fig.9 are read serially by changing the word line address. There are four bits per word for parallel sensing simulation. The peak value of $I_{read}$ for parallel (black solid line) and series sensing (red dotted line) are respectively

~18.4 µA and ~35.6 µA for reading the MTJ in 'P' or '0' state. The dynamic power of reading operation $P_{dynamic}$ can be calculated with Eq.7 and we find that ~8.2% and ~25.7% power could be saved for reading '0' and '1' with the parallel sensing. This asymmetry of power saving is mainly due to the different influence of sneak currents on $R_P$ and $R_{AP}$. According to Ohm's law, a big resistance suffers from higher sneak currents than a small resistance. The mean power saving of parallel data sensing is then ~17%. Note that this simulation is based on a word composed of $N= 4$ bits and the power saving ratio can become more important for a bigger $N$.

The calculation of dynamic power for data programming is more complex as it depends firstly on the STT-MRAM cell. According to the theoretical model of STT dynamic in PMA-MTJ, described in Eq. 6, there is a tradeoff between $I_{write}$ and $\tau$ towards different applications. The switching energy per cell can be thus varied from ~0.6 to ~2 pJ [22] and it cannot be optimized by peripheral circuit design.

As mentioned in subsection II.C, there are sneak currents in the cross-point array due to the floating points in the bit lines (see also Fig.8). Series programming suffers from additional power and the total required write current $I_{write\_final}$ for one STT-MRAM cell programming can be calculated by Eq.8. It is generated by the bit line current source at the right side of cross-point architecture shown in Fig.4.

$$I_{write\_final} = I_{write} + \sum_{i=0}^{M/2} I_{sneak\_i} \quad (8)$$

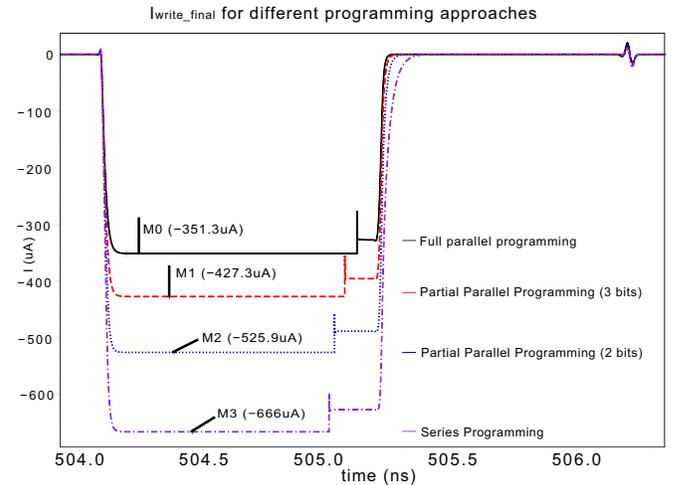

Fig. 14. $I_{write\_final}$ value during different programming approaches: Full parallel programming, partial parallel programming and series bit programming.

The transient simulation shown in Fig.14 can confirm this assumption. By programming one bit in different approaches, we can obtain different $I_{write\_final}$ values. As it is programmed in parallel with all the other bits of the same word, no sneak currents have been observed and $I_{write\_final} = I_{write} = $ ~351 µA (black solid line), which is expected to switch a MTJ in ~1 ns. Sneak currents appear, ~76 µA even though there is only one of the bit lines becomes "floating" and the other three bits are programmed at the same time (red dashed line). As the STT-MRAM cell is programmed bit by bit serially (purple dot-dashed line), $I_{write\_final}$ can be increased up to ~666 µA as there are three "floating" bit lines. This means ~90% power overhead compared with parallel bit programming and it can be increased linearly with a bigger *N*.

One of the most efficient solutions to partially overcome this high power issue of series bit programming is self-enable switching approach [28], which can decrease significantly the number of bit switching operation and minimize as a result the power consumption. For example, if we program a binary data "1010" to a word with previous storage "1011", only one write operation is required. The dynamic switching power is as low as ~1.3 pJ despite the presence of sneak currents, ~46% lower than that of parallel programming ~2.8 pJ.

Based on the above performance analysis, we can conclude that parallel sensing is much better than series sensing in terms of area, power and speed. However, the writing approach would depend greatly on the addressing applications. For embedded solutions, parallel bit programming could be better as it provides fast speed and low power. Nevertheless, series programming could be more suitable for standalone memories, which require high density and provide more power budget.

V. CONCLUSION AND PERSPECTIVE

In this paper, we presented a design of cross-point architecture for STT-MRAM including the program/read methods and the design techniques dedicated to avoid the disturbance influence of sneak currents. Mixed MTJ/CMOS simulation based on 65 nm technology node and analytical calculation were performed to demonstrate its extremely small cell area down to ~1.75 $F^2$ and high data access speed up to some nanosecond per word. We believe that this new cross-point architecture can overcome completely the limitations of 1T+1MTJ structure and allows STT-MRAM to be used as a universal memory in digital computing system.

This cross-point architecture can be extended also to other bi-terminal nanoscale memristive devices [30-33]. If a voltage supply instead of a current is applied to switch the state of these nano-devices like titanium dioxide (TiO2) memristor, word parallel programming approach can be achieved without extra area cost and additional power due to the sneak currents. This helps to increase the data access speed and make up the relatively slow speed of voltage-driven nano-devices.

Unlike 1T+1MTJ structure, cross-point architecture STT-MRAM does not benefit from the matured peripheral circuits of DRAM, and then requires much more R&D effort despite its advantageous cell area and data access speed. For instance, the error correction approach dedicated to improve the reliability of this architecture is under investigation in our lab.

ACKNOWLEDGMENT

The authors wish to acknowledge support the French national projects NANOINNOV-SPIN, PEPS-NVCPU, ANR-MARS and Nano2012 project with STMicroelectronics.

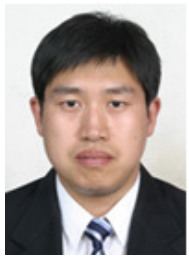

**Weisheng ZHAO** (M'06) was born in Yantai, China, in 1980. He received the M.S.c in Electrical Engineering from ENSEEIHT engineering school, Toulouse, France, in 2004 and the Ph.D degree in physics from the University of Paris-Sud 11, France in 2007. From 2004 to 2008, he investigated Spintronic devices based logic circuits and designed a prototype for hybrid Spintronic/CMOS (90nm) chip in cooperation with STMicroelectronics and French Atomic Agency (CEA). From 2008 to 2009, he was with embedded computing laboratory at CEA and his work included the functional model development and neuromorphic computing architecture design based on nanodevices. Since 2009, he joined the CNRS as a tenured research scientist and his interest includes the hybrid integration of nanodevices with CMOS circuit and new non-volatile memory (40nm technology node and below) like MRAM circuit and architecture design. Weisheng has authored or co-authored more than 40 scientific papers; he is also the principal inventor of 4 international patents.

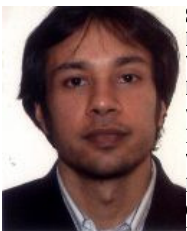

**Sumanta Chaudhuri** received his B.Tech degree in Electronics & Communication Engineering from NIT, Warangal, India in 2000, and PhD. in Electrical Engineering from ENST, Paris, France in 2009. He worked as a Research Engineer in CDoT, Bangalore, India from 2000 to 2004. and was in pursuit of his PhD. Thesis with CNRS and ENST, Paris from 2005 to 2008. During years 2009 and 2010 he was working on hybrid CMOS/MTJ logic circuits in the NanoSpintronics group, IEF, Orsay. His research interests include Reconfigurable Computing, Asynchronous circuits, Cryptography and in a broader sense: physical implementation of computing & communication networks. Currently he is working on variation-aware design as a research associate in circuits & systems group, Imperial College, London.

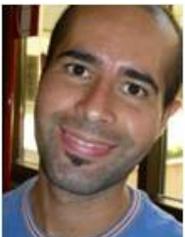

**Celso Accoto** was born in Lecce, Italy in 1979. He received the B.S. degree in Information Engineering from Univ. Lecce in 2007. He is currently studying in an international Master of Science in Micro and Nanotechnologies for Integrated Systems, held by Politecnico di Torino, INP – Grenoble and EPFL. From 2006 to 2010, he was working at Infineon Technologies Development Centre in Padova, Italy, in the Automotive Division. His work included both the design at transistor level of analog blocks and laboratory characterization /validation. In 2011 he joined in IEF, Univ. Paris-Sud, France as an internship student. Here he has designed and simulated the peripheral circuits for Cross-Point MRAM.

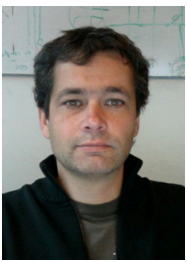

**Jacques-Olivier KLEIN** (M'90) was born in France in 1967. He received the Ph.D. degree and the Habilitation in electronic engineering from the University of Paris-Sud, Orsay, France, in 1995 and 2009 respectively. He is currently professor at *Univ. Paris-Sud*, Orsay in *IEF*, where he leads the nanocomputing research group focusing on the architecture of circuits and systems based on emerging nanocomponents in the field of nanomagnetism and bio-inspired nanonoelectronics. He teaches embedded system design in *Institut Universitaire de Technologie de Cachan*. J.-O. Klein is author of 70 technical papers including 7 invited communications. He served on the conference program Committee like DTIS and GLSVLSI, and he served as reviewer for International Journal of Reconfigurable Computing, IEEE TransMag, Solid State Electronics and conferences. He coordinated the project ANR-PANINI fund by the French Research Agency and he leads, with Cristell Maneux (IMS), the topic "Emerging Technologies" of the Research Group dedicated to system on Chip and System in Package (CNRS GDR SoC-SiP).

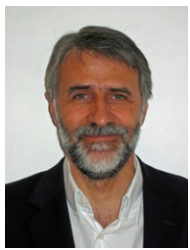

**Claude CHAPPERT** received his "Docteur d'Etat" Diploma in 1985 from Université Paris-Sud, after graduating from the "Ecole Normale Supérieure de Saint Cloud". He is now Research Director at CNRS, with over 30 years experience in research on magnetic ultrathin films and nanostructures, and their applications to ultrahigh density recording. One year was spent as visiting scientist at the IBM Almaden Research Center, San José, USA. He then started a research group on *"Nanospintronics"* within Institut d'Electronique Fondamentale (www.ief.u-psud.fr) of Université Paris-Sud and CNRS. His major interests have been on perpendicular interface anisotropy materials, oscillating interlayer interaction, magnetization reversal in ultrathin films and dot arrays, ion irradiation patterning of magnetic materials, and now spin transfer induced GHz magnetization dynamics of MRAM cells and magnetic logic circuits. He has co-authored more than 250 papers, co-holds 6 patents, and was awarded in 2000 the Silver Medal of CNRS for his research achievements. From 2005 to 2011, Claude Chappert has been in charge of the Spin Electronics Division of the "Centre de Competences en Nanosciences" of Ile-de-France. In January 2010 he has taken the position of Director of IEF.

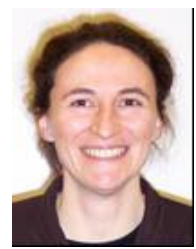

**Pascale Mazoyer** was born in Nîmes, France. She received the M.S. degree in Fundamentals of Physic in 1990 and the PhD degree in physics from Joseph Fourier University in 1994. Her study within CEA LETI was on non-volatile memory (NVM) reliability. Then she has join STMicroelectronics in 1994 on technological process developments addressing successively bipolar, CMOS and embedded memories on advanced technological platforms. She is a cofounder of the international memory workshop (IMW). Today she has joined the IP & Licensing department of STMicroelectronics.